\documentclass{article}

\usepackage{times}
\usepackage{graphicx} 
\usepackage{subfigure}
\usepackage{natbib}
\usepackage{hyperref}
\usepackage{multirow}

\usepackage[accepted]{grcon}

\grcontitlerunning{Demodulation demonstration using the LightCube CubeSat}

\begin{document}

\twocolumn[

\grcontitle{Demodulation demonstration using the LightCube CubeSat}

\grconauthor{Lindsay M. Berkhout}{lmberkhout@asu.edu}
\grconaddress{School of Earth and Space Exploration \\ Arizona State University, Tempe, AZ, USA}
            
\grconauthor{Christopher McCormick}{rommac100@gmail.com}
\grconaddress{Ira A. Fulton Schools of Engineering \\ Arizona State University, Tempe, AZ, USA}

\grconauthor{Daniel C. Jacobs}{daniel.c.jacobs@asu.edu}
\grconaddress{School of Earth and Space Exploration \\ Arizona State University, Tempe, AZ, USA}

\grconauthor{Jaime Sanchez de la Vega}{jaime.a.sanchezdelavega@gmail.com}
\grconaddress{Ira A. Fulton Schools of Engineering \\ Arizona State University, Tempe, AZ, USA}

]

\vskip 0.5in

\begin{abstract}
LightCube is a 1U educational CubeSat which had the goal of connecting the public with space by producing a flash visible to the naked eye on command by a public user. The spacecraft could be triggered via HAM radio communications by those with an amateur license. LightCube is commanded with a DTMF sequence, and reports telemetry using RTTY, an AFSK modulation scheme and is decoded with a custom GNURadio-companion flowgraph. Several radio applications were written, including a from-scratch decoder written for educational purposes and one optimized to be compatible with the SatNOGS environment. Lightcube deployed from the International Space Station on April 24th 2023 and operated for 24 hours before suffering a battery failure. During this time it was tracked by many amateurs around the world with observations reported to the SatNOGs database. Audio observations of the beacons were subsequently decoded by the student team and by amateurs. Having received many observations from around the world, the team has been able to reconstruct the sequence of events leading to loss of communications.  
\end{abstract}

\section{Introduction}\label{Introduction}

The LightCube 1U CubeSat was designed to connect the public with space operations by offering direct interaction with a Low Earth Orbit (LEO) satellite. Additionally, the design and operation of the satellite was largely a student-based effort. Projects like NASA’s CubeSat Launch initiative (CSLI)\footnote{https://www.nasa.gov/directorates/heo/home/CubeSats\_initiative} provide opportunities for educational institutions to fly small satellites, lowering barriers to student involvement in space-based missions. 

\begin{figure}[!ht]
  \begin{center}
    \centerline{\includegraphics[height=\columnwidth,angle=270]{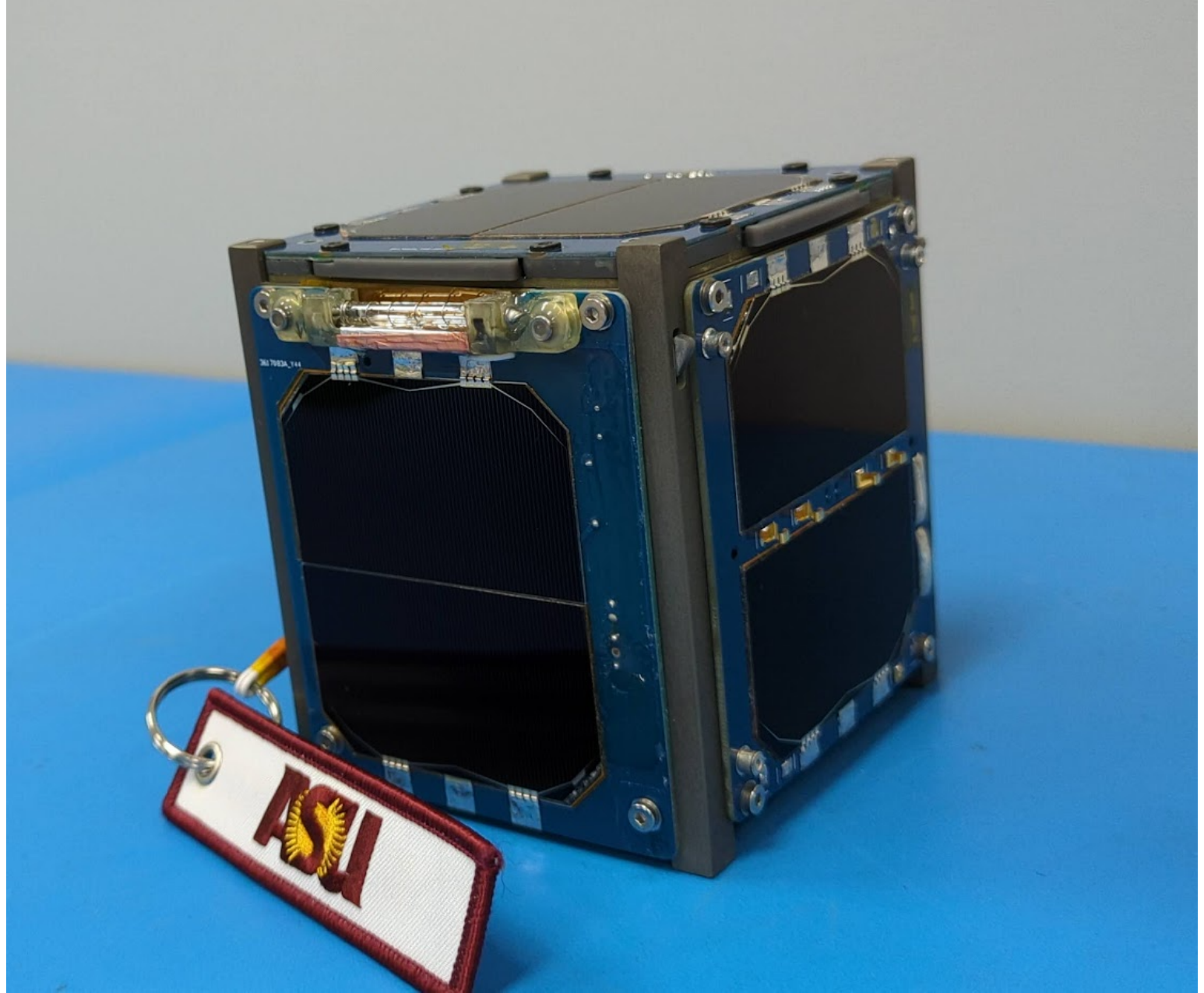}}
    \caption{A benchtop photo of the flight-ready LightCube satellite.}
    \label{realphoto}
  \end{center}
\end{figure}

\newpage

\begin{figure}[ht]
  \begin{center}
    \centerline{\includegraphics[width=\columnwidth]{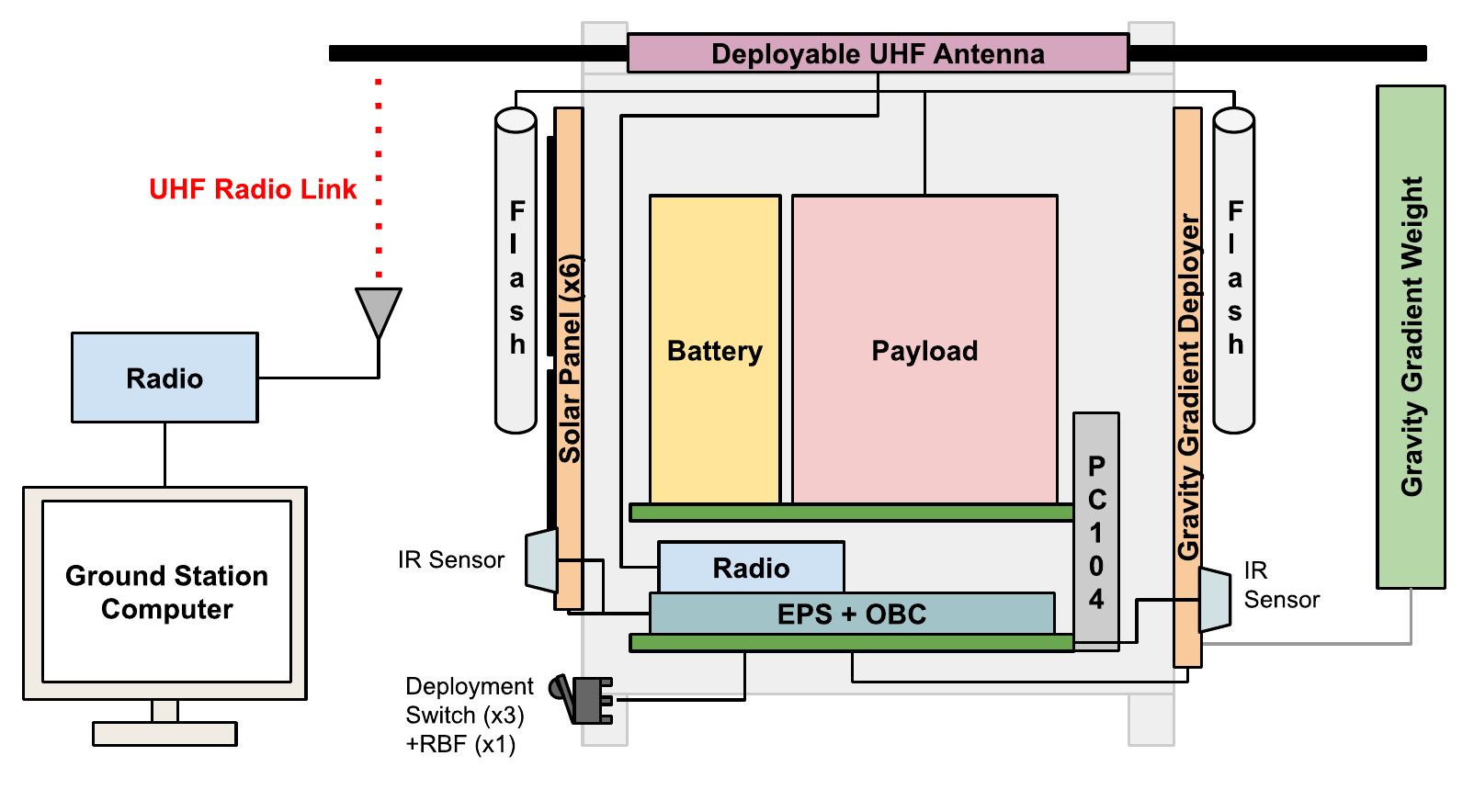}}
    \caption{Block diagram of the satellite. The payload consists of flash bulbs that are naked-eye visible from the ground. Communications between ground station and satellite use UHF frequencies over a deployable antenna.}
    \label{system}
  \end{center}
\end{figure}

The payload consists of a bulb that produces a naked-eye visible flash on user command. In order to command the spacecraft, the only requirement is an amateur HAM radio license. Commands are sent and received through a deployable UHF antenna. The real CubeSat is included in Figure \ref{realphoto}, and a block diagram of the satellite is include in Figure \ref{system}.

During its life the satellite was operated almost exclusively through the SatNOGS amateur satellite network \cite{satnogs} and decoded using GNUradio. This paper provides technical detail on flight and ground side radio systems.

Section~\ref{sec:comms} describes the communications protocols for uplink commands and downlink telemetry. Section~\ref{sec:groundstations} describes the groundstation setups used for received telemetry, and a description of the GNURadio based decoding. Section~\ref{sec:data} describes the parsed telemetry received over the approximately 24 hours of operation.

\section{Spacecraft Communication} \label{sec:comms}
The spacecraft transponder operates in the amateur Ultra High Frequency (UHF) band,  using a deployable UHF antenna to transmit and receive from groundstations. The antenna is a circularly polarized dual dipole configuration.

The spacecraft accepts commands through a Dual Tone Multi-Frequency (DTMF) sequence transmitted by a ground user, and Downlink modes include Radioteletype (RTTY) with Audio Frequency Shift Keying (AFSK) for telemetry and DTMF tone sequences for command confirmation. The transceiver operates at 437.175 MHz in the amateur band under coordination by the International Amateur Radio Union (IARU) with call sign KJ7TZG. The radio itself is a DRA818U-UHF Band Voice Transceiver Module, and a CMX865A modem provides both the DTMF and RTTY (AFSK) encoding and decoding of the output audio. 

 A summary of the communications protocol information is available in table~\ref{table:links}.

\begin{table}[!ht]
\caption{Scheme, protocol, and carrier information for Lightcube.}
\begin{center}
\begin{tabular}{ |c|c|c| } 
 \hline
  & Scheme  & Carrier \\ 
 \hline
 Downlink & RTTY(AFSK) & 437.175 MHz \\ 
 Uplink & DTMF & 437.175 MHz\\ 
 \hline
\end{tabular}
\end{center}
\label{table:links}
\end{table}

\subsection{Uplink Commands}
The spacecraft payload is commanded using a sequence of DTMF codes. The Xenon flash tubes have an 8 $\mu s$ flash length, and flashes are limited to every 30 seconds to allow the payload to reset. The current sequences are unpublished, as the team was unable to test the payload in-situ before communications failure.  The payload flash can be enabled or disabled by the communications team to allow amateur users to signal the spacecraft. Should a flash command be issued, it will be logged and publicly available, enabling a historical recording of commands. Table~\ref{table:dtmf} displays the frequency mapping for DTMF tones. 

\begin{table}[!ht]
\caption{Tone encoding used for Dual Tone Multi-Frequency (DTMF) signaling}
\vspace{2mm}
\label{table:dtmf}
\begin{tabular}{ |c|c|c|c|c| } 
 \hline
  Tones & 1209 Hz & 1336 Hz & 1477 Hz & 1633 Hz \\ 
 \hline
  697 Hz & 1 & 2 & 3 & A \\ 
  770 Hz & 4 & 5 & 6 & B \\ 
  852 Hz & 7 & 8 & 9 & C \\ 
  941 Hz & * & 0 & \# & D \\ 
 \hline
\end{tabular}
\end{table}

\subsection{Downlink Telemetry}

A single downlink transmission can be one of three types: a full telemetry packet, an orientation packet, and a `heartbeat' packet. During operation, full telemetry was sent every one minute, and the heartbeat and orientation packets were sent every 30 seconds. Telemetry is encoded with RTTY (AFSK) at 300 Baud. Full packet definitions are available online\footnote{https://interplanetarylab.github.io/missions/Lightcube}, and include satellite health information such as solar panel metrics and flash bulb status. 

\begin{table}[!ht]
\caption{Baud rate and mark/space frequency information for the Bell 103 protocol used for downlink telemetry}
\vspace{2mm}
\label{table:Bell103}
\begin{tabular}{ |c|c|c|c|c| } 
 \hline
  Bell 103 FSK Mode & Min & Typ. & Max & Units \\ 
 \hline
 Baud rate &  & 300 & & \\ 
 Mark (logical 1)  & 2222 & 2225 & 2228 & Hz \\ 
 Space (logical 0)  & 2022 & 2025 & 2028 & Hz \\ 
 \hline
\end{tabular}
\end{table}

\begin{figure*}[!ht]
  \begin{center}
    \centerline{\includegraphics[width=1.8\columnwidth]{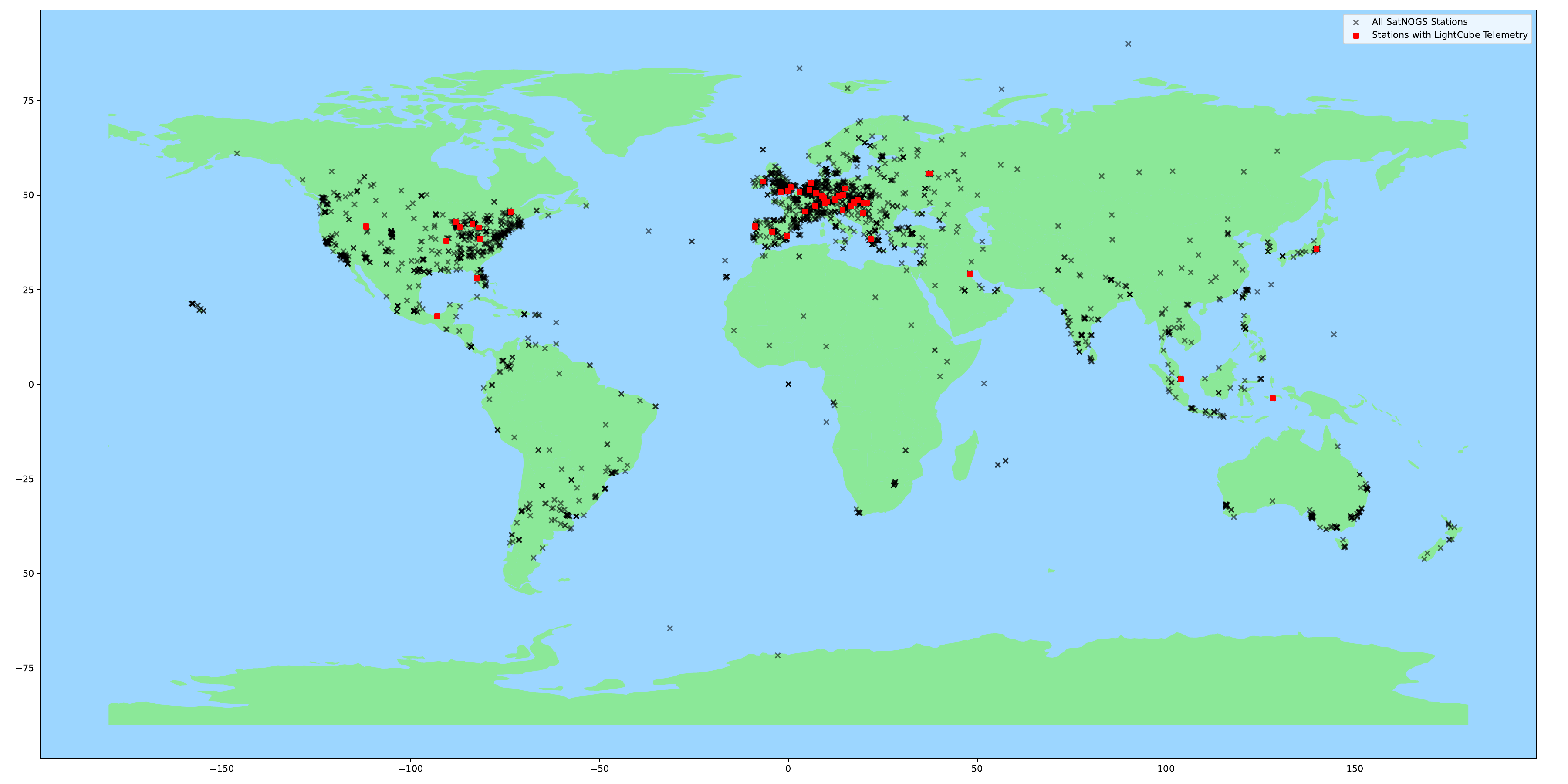}}
    \caption{Locations of all SatNOGS registered stations, indicated with black 'x' markers. Square red markers show stations with received LightCube packets.}
    \label{stations}
  \end{center}
\end{figure*}

\section{Groundstation Communications} \label{sec:groundstations}

\subsection{SatNOGS}
Support from the Libre Space Foundation Satellite Networked Open Ground Stations (SatNOGS) project was critical to operations. SatNOGS provides a network of amateur groundstations around the world. SatNOGS provides a guide for amateur radio operators to set up receiver stations. These stations are based on software defined radios which connect through a standardized program back to a central server. On this server remote users can schedule observations of satellites and inspect the results.

SatNOGS supports many amateur satellites with observations and decoding. Satellite teams can integrate their instrument into the SatNOGS architecture by adding relevant information (modulation scheme, carrier frequency, etc.) to the database of supported satellites and providing a GNURadio based decoder to parse the recieved packets.

Figure~\ref{stations} shows the locations of all registered SatNOGS stations. Stations that received successfully demodulated LightCube beacons are highlighted. 

\begin{figure}[!ht]
  \begin{center}
    \centerline{\includegraphics[width=\columnwidth]{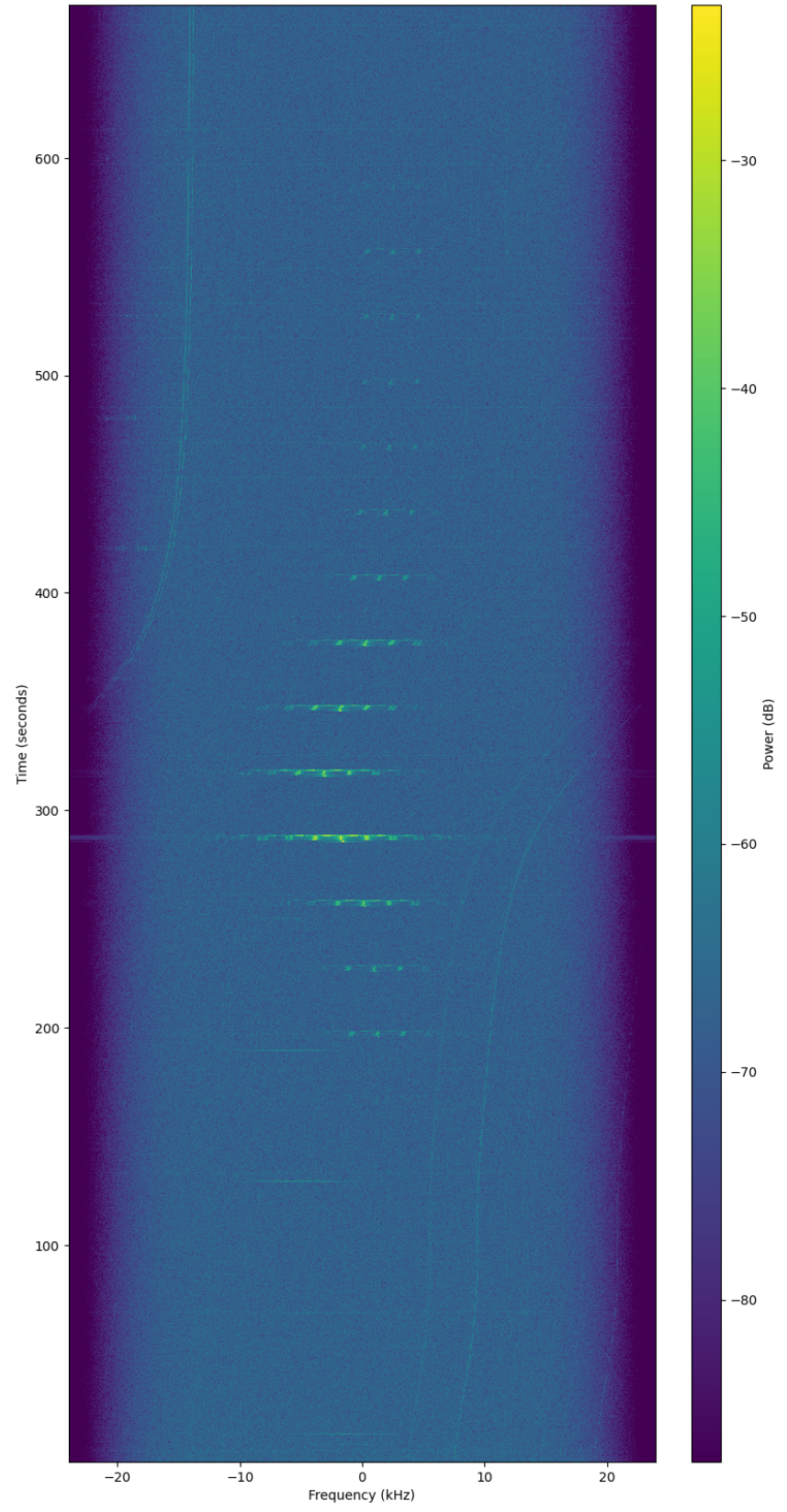}}
    \caption{Observation by Fredy Damkalis on station owned by Mike Rupprecht. The SatNOGS network demodulates the FM carrier and outputs an audio file as well as a figure showing the baseband signal, included here. For LightCube, this audio output needs to be further demodulated for the AFSK scheme.}
    \label{satnogs}
  \end{center}
\end{figure}

Figure~\ref{satnogs} shows an example high-quality LightCube observation from a SatNOGS station. SatNOGS offers an FM demodulated audio file as a station product, as well as a waterfall plot showing the FM demodulated signal. 

The SatNOGS network also offers direct demodulation of received signals for satellites that provide accompanying GNURadio flowgraphs. The LightCube demodulator itself is not currently integrated into the network, and telemetry outputs are processed by hand from the provided audio files. 

\begin{figure*}[!ht]
  \begin{center}
  \centerline{\includegraphics[width=1.8\columnwidth]{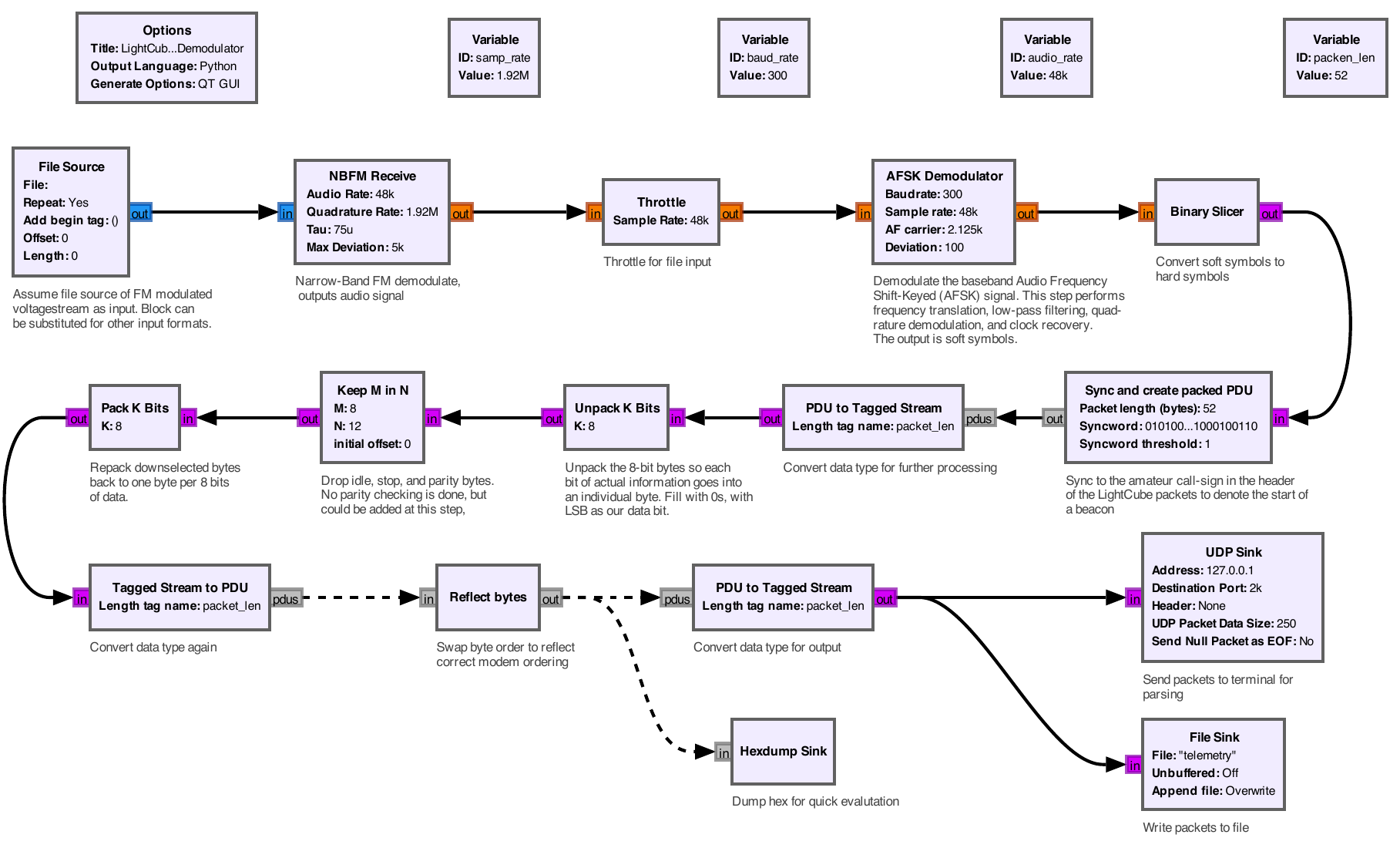}}
    \caption{Annotated GNURadio Companion flowgraph for demodulating LightCube telemetry packets. This version assumes a voltage stream file, but other versions of the flow graph allow for live hardware decoding or post-hoc input of a SatNOGS audio file. To parse (assign bytes to telemetry values), a Python program captured packets from the terminal or output file and printed the values.}
    \label{annotated}
  \end{center}
\end{figure*}

\subsection{ASU Groundstation}
The local ASU Groundstation is connected to the SatNOGS network, and additionally uses GQRX\footnote{https://gqrx.dk/} for some applications.  The GPredict software\footnote{http://gpredict.oz9aec.net/} is used for tracking and doppler correction.

The hardware consists of two 70cm Yagi Antennas, with a Yaesu G5500 Rotator and Kuhne LNA. A few different radios are available, including software defined radios such as the RTL-SDR and HackRF-One, and a traditional ICOM-9100 radio.

\subsection{Decoding with GNURadio}
Decoding of LightCube beacons is enabled by a GNURadio-companion flowgraph. An annotated version of the flowgraph is included in Figure~\ref{annotated}. The version in this figure expects a raw voltage steam from a file, but other versions of are available to allow for decodes from live hardware streaming or SatNOGS audio files. 

Received data must first be FM demodulated to extract the audio signal from the FM modulated carrier at 437.175 MHz. This step is performed within the flowgraph in the case of a hardware input or voltage stream file by the NBFM block, which then outputs the baseband audio data. However, the SatNOGS network outputs FM demodulated audio data already, so this step is bypassed for the SatNOGS flowgraph version. 

A throttle block can be enabled or bypassed depending on the input source. Hardware inputs do not need the throttle, while it must be included when using a file input as issues with sample rates will occur otherwise.

Following the throttle block, soft symbols are extracted from the audio data. This step also offers options depending on user-choice. While the stream is now demodulated from the carrier wave, the audio signal itself must be translated to baseband and low pass filtered. Then the frequency shift keyed signal is quadrature demodulated and clock recovery is performed. These steps are performed using individual GNURadio blocks in one version of the flowgraph for educational purposes, as the user can follow the demodulation process step-by-step. For a simplified flowgraph, there is an option to use a pre-built demodulator block.

The annoted version included in Figure~\ref{annotated} shows the use of this pre-built AFSK demodulator block from the gr-satellites package\footnote{https://github.com/daniestevez/gr-satellites} to perform the soft symbol extraction steps. Gr-satellites is an out-of-tree GNURadio module providing support blocks for satellite communications. The project supports many amateur satellites and modulation schemes.   

The soft symbols are converted to hard symbols by a binary slicer, which is fed to the `sync and create packed PDU' block provided by the gr-satellites package. This block searches for a defined sequence to denote the start of a packet. For LightCube, the sync searches for the amateur radio call-sign included in the packet header. The PDU is converted back to a tagged stream to perform some data manipulation. The modem produces two `idle' bits, as well as a stop and parity bit. To facilitate easy parsing, these bits are removed and the bytes reversed to reflect the modem ordering. 

The bytes now represent a complete and parse-able telemetry packet. The stream is dumped to debug and hexdump blocks within GNURadio-companion for a quick view, and simultaneously written to a file and sent over UDP to the local user terminal. The output packet is parsed to assign specific telemetry parameters to their corresponding bytes.

\section{Received Telemetry} \label{sec:data}
The SatNOGS network received the first confirmed beacon from LightCube at 13:31:26 UTC on April 24th, 2023. Beacons were consistently recorded until 08:55:22 UTC on April 25th, 2023. The received beacons appear to cover around 9 orbits. Using decoded packets from SatNOGS and the ASU ground station, the communications team was able to reconstruct the likely series of events leading to loss of communication.

Reported values for the satellite battery temperature suggested unexpected variability. The variability was tracked to a software mistake which left the battery heater turned off in a pre-launch testing configuration. 

\begin{figure}[!ht]
  \begin{center}
    \centerline{\includegraphics[width=\columnwidth]{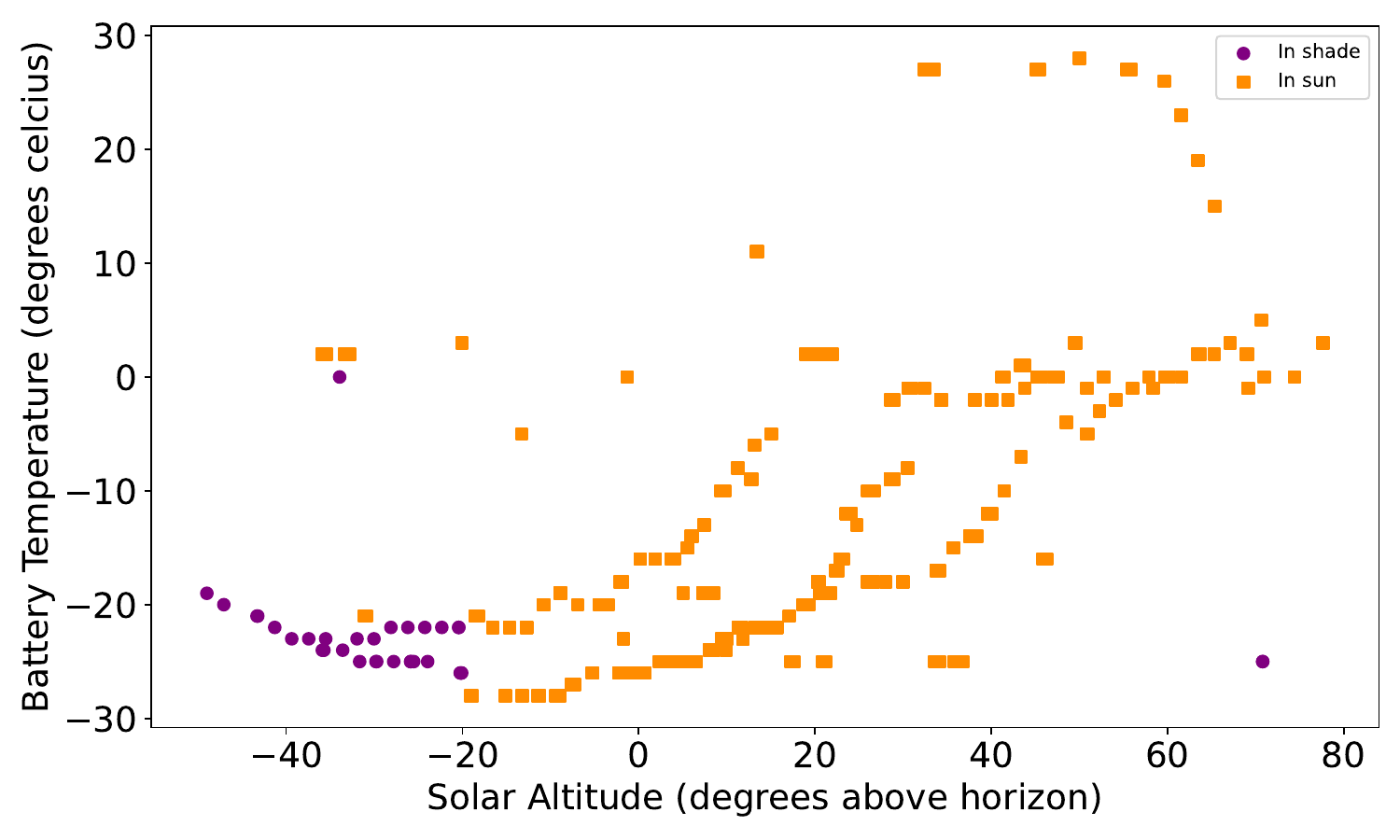}}
    \caption{Solar altitude at satellite location vs. battery temperature reported from satellite telemetry. The data covers telemetry from 9 orbits. Purple circle data was were the satellite was determined to be `in-shade' by the Skyfield \cite{skyfield} package, and the orange square data was determined to be `in-sun'. Battery temperature trends appear to largely follow Solar altitude and `in-sun' determinations.}
    \label{data}
  \end{center}
\end{figure}

This left battery temperature at the mercy of natural solar heating, and the temperature therefore swings with sun/shade exposure. Often, the batteries fell below the recommended operating temperature of  0$^{\circ}$C. Battery temperature as a function of local Solar angle, shown in Figure~\ref{data} gives some idea of how solar input might have driven battery heating. Beacons are color/shape coded by an `in-sun' or `in-shade' determination from the Skyfield \cite{skyfield} package. Trends largely suggest that the battery temperature experienced peaks and troughs correlated with Solar irradiance. 

Based on the temperatures reported well below the recommended battery operating minimum of  0$^{\circ}$C , it is not likely we will receive more beacons from the satellite. However, a potential remains for communication to be restored during times when the satellite has been warmed by the sun for extended periods as the orbit precesses and solar angles increase.  

\newpage

\section{Conclusion}
Using the large coverage and reliable stations of the SatNOGS network and GNURadio based software tools, telemetry was obtained and demodulated from the LightCube CubeSat during a very short window when the satellite operated. Battery temperature data from the telemetry suggests that the hardware has fallen below allowed temperature ranges, and the CubeSat ceased reporting telemetry shortly under 24 hours after launch. Some potential exists for beacons to resume due to natural heating from high Solar intensity orbits, but based on the long lapse in communications it is not likely. 

Much was learned from the process of launching the satellite, and lessons learned will be applied to future educational satellite projects. LightCube was developed and operated largely by students, offering early involvement in real space projects. 

\section*{Software}
This work was enabled by a number of software packages, including the geopandas package\footnote{https://github. com/geopandas/geopandas} for station plotting, the Astropy \cite{astropy} and Skyfield packages for satellite orbit information, and the GNURadio project. We would also like to acknowledge the use of the out-of-tree GNURadio-companion modules provided by SatNOGS and gr-satellites. 

The GNURadio based decoders for this project are availble at \url{https://github.com/ASU-cubesat/lightcube_telemetry}.

\section*{Acknowledgements}
LightCube is supported by funding from the  Interplanetary Lab at Arizona State University (ASU), Vega Space Systems, and ASU Alumni. 

We thank the amateur community involved in the SatNOGS project for enabling high duty cycle observations of the LightCube satellite. 

LMB acknowledges that this material is based upon work supported by a National Science Foundation Graduate Research Fellowship under Grant No. 2233001.

\bibliography{Lightcube}
\bibliographystyle{grcon}

\end{document}